\documentstyle[aps,pra]{revtex}
\tightenlines
\newcommand{\be}{\begin{equation}}
\newcommand{\ee}{\end{equation}}
\newcommand{\ben}{\begin{eqnarray}}
\newcommand{\een}{\end{eqnarray}}
\newcommand{\bF}{\begin{figure}}
\newcommand{\eF}{\end{figure}}
\title{On the Incompatibility of Standard Quantum
Mechanics and the de Broglie-Bohm Theory}
\author{Partha Ghose}
\address{S. N. Bose National Centre for Basic Sciences,
Block JD, Sector III, Salt Lake, Kolkata 700 098}

\begin{document}
\maketitle

\begin{abstract} It is shown that the de
Broglie-Bohm quantum theory of multi-particle systems is
incompatible with the standard quantum theory of such systems
unless the former is ergodic. A realistic experiment is suggested
to distinguish between the two theories.
\end{abstract}


\section{Introduction}

Quantum mechanics has been plagued by interpretational problems
since its inception. This is rooted in the measurement problem
which has defied all attempts at a satisfactory solution
\cite{wheeler}. Once the wavefunction is assumed to contain
complete information of a system, the measurement problem is an
inescapable consequence of its linear, unitary Schr\"odinger
evolution. This inevitably implies the lack of (a) determinism at
a fundamental level, (b) of Einstein-Podolsky-Rosen type of
reality \cite{epr} and (c) Einstein-Bell locality \cite{bell}, all
of which raise deep interpretational and philosophical problems.
One possible way out of these problems is to give up the
assumption that the wavefunction contains complete information of
the system and introduce hidden variables to restore determinism
and reality at a deeper level. It has been shown, however, that
all {\it local} hidden variable theories and all {\it
non-contextual} hidden variable theories (theories in which the
experimental set-up or context in which a measurement is made
plays no role) are incompatible with quantum mechanics
\cite{kochen}. The only hidden variable theory that has so far
survived all incompatibility theorems and experimental
falsification is the de Broglie-Bohm theory (dBB)
\cite{bohm},\cite{Holland} which is both contextual and non-local.

The purpose of this paper is to demonstrate that even dBB is
incompatible with SQT for multiparticle systems that are
non-ergodic in dBB, an aspect that has not been analyzed before.
Seen in this light, dBB is not merely an interpretation of SQT as
was originally intended by Bohm but a different physical theory
with different predictions in special circumstances that have not
been tested so far. A realistic experiment will be described which
can distinguish between dBB and SQT. This experiment will once and
for all settle the dispute between critics of (i) determinism and
(ii) EPR type reality and their opponents.

A simple and brief discussion of ergodicity in classical and
quantum mechanics will be found in sections $\rm{II}$ and
$\rm{III}$. Section $\rm{IV}$ contains a discussion of joint
detection probabilities that are crucial in distinguishing between
dBB and SQT, and the theory behind a realistic experiment to do so
is decribed in section $\rm{V}$.

\section{Ergodicity in classical mechanics}

I will begin by giving a simple example from classical mechanics
to introduce the nomenclature and the basic features of ergodicity
that will be useful for my purpose. Consider the familiar
classical system of two identical simple pendulums of length $l_1
= l_2 = 1$ and mass $m_1 = m_2 = 1$ connected by a weightless
spring whose length $\ell$ is equal to the distance between the
points of suspension. If $q_1$ and $q_2$ denote the angles of
inclination of the pendulums, then for small oscillations the
kinetic energy $T = \frac{1}{2}(\dot{q}_1^2 + \dot{q}_2^2)$ and
the potential energy $U = \frac{1}{2}(q_1^2 + q_2^2 + \alpha (q_1
- q_2)^2)$ where $\alpha (q_1 - q_2)^2$ is the potential energy of
the elastic spring. Now define the normal coordinates

\begin{equation}
Q_1 = \frac{q_1 + q_2}{\sqrt{2}}\,\,\,\,\,\, {\rm and}\,\,\,\,\,\,
Q_2 = \frac{q_1 - q_2}{\sqrt{2}}
\end{equation}
Then,

\begin{equation}
T = \frac{1}{2}(\dot{Q}_1^2 + \dot{Q}_2^2)\,\,\,\,\,\, {\rm
and}\,\,\,\,\,\, U = \frac{1}{2}(\omega_1^2 Q_1^2 + \omega_2^2
Q_2^2)
\end{equation} where $\omega_1 = 1$ and $\omega_2 = \sqrt{1 + 2
\alpha}$. So, the characteristic oscillations are: \vskip 0.2in
\noindent 1. $Q_2 = 0,\,\,\,\,\,{\rm i.e.,}\,\,\,\, q_1 = q_2$ and
the two pendulums oscillate in phase with the original frequency
$\omega_1 = 1$, or

\noindent 2. $Q_1 = 0,\,\,\,\,\,{\rm i.e.}\,\,\,\, q_1 = - q_2$
and the two pendulums oscillate with opposite phase with the
increased frequency $\omega_2 > 1$. \vskip 0.2in \noindent The
smooth phase-space manifold $M$ on which the motion occurs is the
torus $T^2= S^1 \times S^1$, and the orbits are closed curves on
this torus. This shows that the system is non-ergodic. What that
means is that the orbits are not everywhere dense on the torus, or
intuitively, the orbits do not cover the entire available phase
space (the energy surface) even if one waits infinitely long
\cite{Arnold1}. If one regards the system as a two-dimensional
oscillator rather than two one-dimensional ones that are coupled,
the system will still be non-ergodic provided $\omega_1/\omega_2$
is a rational number. If $\omega_1/\omega_2$ is irrational, the
system will be ergodic, i.e., the orbits will not be closed curves
and will be everywhere dense on the torus.

The non-ergodic character of a dynamical system results in a
difference between the space and time means of its dynamical
variables. The space and time means of a complex valued function
$F$ on $M$ are defined by \ben \bar{F}&=& \int_M
F(q,p)\,\rho(q,p)\, dq\, dp,\,\,\,\,\int \rho(q,p)\, dq\, dp = 1
\\ F^* &=& {\rm lim}_{N \rightarrow \infty} \frac{1}{N}
\sum_{n = 0}^{N - 1} F (\phi_t^n
q)\label{eq:timeav}\een where $q$ and $p$ stand for the set of
coordinates and momenta, $\rho(q,p)\, dq\, dp$ for the invariant
measure in phase space (Liouville's theorem), and $\phi_t: M
\rightarrow M$ a one parameter (time) group of measure preserving
diffeomorphisms. There are fundamental theorems in ergodic theory
which state that the space and time means of every complex valued
function $F$ on $M$ exist and will be identical if the system is
ergodic, and cannot be the same if the system is non-ergodic
\cite{Arnold2}.

\section{Ergodicity in quantum mechanics}

The same system of oscillators is described in standard quantum
theory (SQT) by the two-particle Schr\"{o}dinger equation

\begin{equation}
i \hbar \frac{\partial \psi (Q_1, Q_2)}{\partial t} = [ -
\frac{\hbar^2}{2} \partial_{Q_1}^2 - \frac{\hbar^2}{2}
\partial_{Q_2}^2 + \frac{1}{2} \omega_1^2 Q_1^2 + \frac{1}{2} \omega_2^2
Q_2^2 ]\psi (Q_1, Q_2)
\end{equation}
\begin{equation}
i \hbar \frac{\partial \psi (Q_1, Q_2)}{\partial t} = [ -
\frac{\hbar^2}{2} \partial_{Q_1}^2 - \frac{\hbar^2}{2}
\partial_{Q_2}^2 + \frac{1}{2} \omega_1^2 Q_1^2 + \frac{1}{2} \omega_2^2
Q_2^2 ]\psi (Q_1, Q_2)
\end{equation}
One can then construct two non-dispersive wave-packets oscillating
about $Q_1 = a_1$ and $Q_2 = -a_2$ \cite{Holland}. Let

\begin{eqnarray}
\psi_A (Q_1, t) &=& (\omega_1/\pi \hbar)^{1/4} {\rm exp}\,\,\{ - (
\omega_1 /2 \hbar) (Q_1 - a_1\, {\rm cos}\, \omega_1 t)^2\\\nonumber
&-& (i/2)[ \omega_1 t + (\omega_1 / \hbar)(2 Q_1 a_1\, {\rm sin}\,
\omega_1 t - \frac {1}{2} a_1^2 {\rm sin}\, 2 \omega_1 t)]\}
\label{eq:wp1}
\end{eqnarray}
be the packet initially centred about $Q_1 = a_1$ and

\begin{eqnarray}
\psi_B (Q_2, t) &=& (\omega_2/\pi \hbar)^{1/4} {\rm exp}\,\,\{ - (
\omega_2 /2 \hbar) (Q_2 + a_2\, {\rm cos}\, \omega_2 t)^2\\\nonumber
&-& (i/2)[ \omega_2 t + (\omega_2 / \hbar)( - 2 Q_2 a_2\, {\rm
sin}\, \omega_2 t - \frac {1}{2} a_2^2 {\rm sin}\, 2 \omega_2 t)]\}
\label{eq:wp2}
\end{eqnarray}
the packet initially centred about $Q_2 =-a_2$. The packets oscillate harmonically without
change of shape between the angles $a_1$ and $- a_2$ . The two-particle wavefunction is given
by

\begin{equation}
\psi(Q_1, Q_2, t) = \psi_A(Q_1, t) \psi_B (Q_2, t) = R(Q_1, Q_2, t
)\,{\rm exp}\,\, \frac{i}{\hbar} S(Q_1, Q_2, t)
\label{eq:WF}
\end{equation}
and therefore the phase or action function by

\begin{eqnarray}
S(Q_1, Q_2, t) = &-& \frac{1}{2} \hbar \omega_1 t - \frac{1}{2}
\omega_1 ( 2 Q_1 a_1\, {\rm sin}\, \omega_1 t - \frac{1}{2} a_1^2\,
{\rm sin}\, 2 \omega_1 t)\\\nonumber &-& \frac{1}{2} \hbar
\omega_2 t - \frac{1}{2} \omega_2 ( - 2 Q_2 a_2\, {\rm sin}\, \omega_2
t - \frac{1}{2} a_2^2\, {\rm sin}\, 2 \omega_2 t)
\end{eqnarray}
The Bohmian trajectory equations are therefore

\begin{eqnarray}
P_1 &=& \frac{d Q_1}{d t} = \partial_{Q_1} S(Q_1, Q_2, t) = - \omega_1
a_1\, {\rm sin}\, \omega_1 t\\ P_2 &=& \frac{d Q_2}{d t} = \partial_{Q_2}
S(Q_1, Q_2, t) =  \omega_2 a_2\, {\rm sin}\, \omega_2 t
\label{eq:mom}
\end{eqnarray}
whose solutions are

\begin{eqnarray}
Q_1(t) &=& Q_1(0) + a_1\,\, ({\rm cos}\, \omega_1 t - 1)\\\nonumber Q_2(t)
&=& Q_2(0) - a_2\,\, ({\rm cos}\, \omega_2 t - 1)
\label{eq:tr}
\end{eqnarray}
where $Q_1(0)$ and $Q_2(0)$ are the initial coordinates. If one considers an ensemble
of such oscillators, their centre points are distributed in a gaussian fashion.
The characteristic oscillations are again: \vskip 0.2in \noindent
1. $Q_2(t) = 0$, i.e., $q_1(t) = q_2(t)$, and the two particles
oscillate in phase with the original frequency $\omega_1$ (and
hence with the length $\ell$ of the spring unchanged), or

\noindent 2. $Q_1 (t) = 0$, i.e., $q_1(t) = - q_2(t)$, and the two
particles oscillate out of phase with the increased frequency
$\omega_2$. \vskip 0.2in \noindent

The quantum potentials of the
two oscillators turn out to be \ben Q(1) &=& \frac{1}{2}\hbar
\omega_1 - \frac{1}{2}\omega_1^2 (Q_1(t) - a_1 {\rm cos} \omega_1
t)^2\nonumber\\ Q(2) &=& \frac{1}{2}\hbar \omega_2 -
\frac{1}{2}\omega_2^2 (Q_2(t) + a_2 {\rm cos} \omega_2 t)^2 \een
Thus, $Q(1)$ and $Q(2)$ are constants on the trajectories
(\ref{eq:tr}). Using these results, one obtains
\begin{eqnarray}
\frac{1}{2} (P_1^2(t) + \omega_1^2 Q_1^2(t)\,) + Q(1) &=&
\frac{1}{2}\hbar \omega_1 + \frac{1}{2} \omega_1^2 a_1^2 +
\omega_1^2 (Q_1(0) - a_1) {\rm cos} \omega_1 t \\  \frac{1}{2}
(P_2^2(t) + \omega_2^2 Q_2^2(t)\,) + Q(2) &=& \frac{1}{2}\hbar
\omega_2 + \frac{1}{2}\omega_2^2 a_2^2 + \omega_2^2 (Q_2(0) + a_2)
{\rm cos} \omega_2 t \end{eqnarray}
The motion is still on a torus $T^2$ in each case (1 and 2) with the size of
the torus oscillating in time about a mean value. Since the motion
is periodic, the system must be non-ergodic. Then it follows from
the general theorem in ergodic theory that there must be at least
one observable of the system whose space and time averages are
different.

Having demonstrated that a system in dBB can be non-ergodic, I
will now make use of a von Neuman-Birkhoff-Khinchin type
ergodicity theorem in quantum mechanics which shows that all
systems in SQT are ergodic \cite{toda}. This theorem is well-known
among mathematicians, and therefore I will only sketch a simple
proof for non-degenerate systems. Consider a two-particle quantum
mechanical system with a discrete, non-degenerate energy spectrum.
Let $\Psi (x_1, x_2, t) = {\rm exp}\, (- i H t/\hbar)\,\psi (x_1,
x_2)$ be a normalized solution of the time-dependent
Schr\"{o}dinger equation, and let $\psi (x_1, x_2) = \sum_n c_n
\phi_n (x_1, x_2)$ where $\phi_n (x_1, x_2)$ are a complete set of
orthonormal energy eigenfunctions. Since the average over states
in Hilbert space or wavefunctions does not have any direct
physical interpretation, we follow Ref. \cite{Parry} and consider
the time average of the expectation value in the state $\Psi (x_1,
x_2, t)$ of a hermitian operator $\hat{F}$ that does not commute
with the Hamiltonian :

\begin{eqnarray}
F^* &=& {\rm lim}_{T \rightarrow \infty}\frac{1}{T} \int_0^T dt
\int dx_1 dx_2\, \Psi^*(x_1,x_2,t)\,\hat{F}
\Psi(x_1,x_2,t)\\\nonumber &=& {\rm lim}_{T \rightarrow
\infty}\frac{1}{T} \int_0^T dt \int dx_1 dx_2\, (\,\sum_n \vert
c_n  \vert^2 \phi_n^* (x_1, x_2) \hat{F} \phi_n (x_1, x_2)
\\\nonumber &+& \sum_{n,m} c^*_n c_m \,e^{i (E_n - E_m)t}
\phi^*_n (x_1,x_2)\,\hat{F} \phi_m (x_1,x_2)\,)\\\nonumber &=&
\sum_n \vert c_n \vert^2 \int dx_1 dx_2 \phi_n^* (x_1, x_2)\,
\hat{F}\,\phi_n (x_1, x_2)\\\nonumber &=& Tr (\hat{\rho}
\hat{F})\\\nonumber &\equiv& \bar{F}
\end{eqnarray}
because $E_n \neq E_m$. $\hat{\rho}$ is the reduced density matrix
obtained after time averaging. It is clear in this case that the
limit exists, is unique, non-vanishing and time independent, and
also equals the space average defined with the reduced density
operator corresponding to the weighted average of the eigenvalues
of $\hat{F}$. In the absence of a well defined phase space in SQT,
this is the general criterion of ergodicity in SQT. If $\hat{F}$
commutes with the Hamiltonian, then it is clear from the above
that $F^* = Tr (\rho \hat{F}) = \bar{F} = \langle \hat{F} \rangle$
where $\rho$ is the density matrix of the pure state $\Psi$. This
theorem can be trivially generalized to multi-particle systems.
However, its generalization to systems with a continuous energy
spectrum is non-trivial. The proof is complicated and involves
coarse-graining and random-phase approximations, and the reader is
referred to Ref. \cite{toda} and \cite{Parry} for further details.

Notice that the time averaged reduced density matrix in the proof
of ergodicity given above is the same as the reduced density
matrix introduced by von Neumann through the collapse postulate.
It is generally believed that the measurement process embodied in
the collapse postulate must be instantaneous. An ideal
measurement, on the other hand, involves an accurate measurement
of energy which takes an infinitely long time. This apparent
contradiction was resolved by von Neumann \cite{von} in the
following way. To quote him,
\begin{quote}`` ...what we really need is not that the change of
$t$ be small, but only that it have little effect in the
calculation of the probabilities ... That is, the state $\phi_n$
should be essentially ... a stationary state; or equivalently
...$\phi_n$ an eigenfunction of $H$.''
\end{quote}
This is indeed what has been used in the proof of ergodicity given
above.

To recapitulate, I have so far shown that SQT systems are
necessarily always ergodic but that the corresponding dBB as well
as classical systems are not necessarily so. I will now show how
this can lead to observable differences between dBB and SQT.

\section{Joint detection probabilities}

The demonstration rests on the special ontological status that the
particle position has in dBB. To quote Holland \cite{Holland3},
\begin{quote}
``Although the general theory of measurement entails a disturbance
of the initial wavefunction so that the actual value found for the
particle property is not the preexisting value, there is one
important exception to this rule : ideal measurements of position.
These hold a unique significance in the theory for, while the
initial arbitrary wavefunction is transformed, it condenses around
the current position of the particle and so we are able, in
principle, to infer the premeasurement position as defined by the
causal interpretation."
\end{quote}
This special status can be exploited to design experiments that
can distinguish between dBB and SQT. In this context the joint
detection of two positions is of special interest.
In order to do that one must first define the space average of a
dynamical variable in dBB. Unlike in SQT, it is possible to define
a phase space in dBB, and through it the space average of
dynamical variables. A joint distribution function $f(q,p,t)$ can
be defined in dBB by \cite{Holland1}

\begin{eqnarray}
f(q,p,t) &=& P(\,q(t)\,) \delta(p - \nabla S(q,t))\\
\int f(q,p,t) dq dp &=& 1
\end{eqnarray}
where $P(\,q(t)\,)$ is the real statistical probability density in
dBB that is equivalent to the quantum mechanical probability
density $\vert \psi (q,t) \vert^2$. Take any function $F(q,p)$ on
phase-space. Its space average is defined by

\begin{eqnarray}
\bar{F} &=& \int F(q,p) f(q,p,t) dq dp\nonumber\\ &=& \int F(q,
\nabla S) P(\,q(t)\,) dq\label{eq:spav}
\end{eqnarray}
which is the same as the ensemble average. It must be emphasized that
the premeasurement momentum $p$ defined
in this way will not generally agree with the measured momentum
because momentum does not have the same ontological status as
position in dBB. Nevertheless, dBB does lead to the correct
prediction for momentum consistent with SQT and the uncertainty
relations once the effect of measurement is taken into account
\cite{bohm},\cite{Holland}. This is a consequence of the fact that
dBB is constructed in such a way as always to have the space or ensemble
average of every observable identical with its quantum mechanical
expectation value.

\section{Two-particle interferometer experiment}

Before concluding, I will discuss another system for which dBB and
SQT are incompatible and which can be used to distinguish between
them in practice. Consider a source of two momentum correlated
identical particles of mass $m$ (in the sense of the original EPR paper)
(described by wave packets) set up
in such a fashion that they pass simultaneously through two point
slits $A$ and $B$ situated on the $y$ axis and separated by a
distance $2 a$. Let only one pair of packets pass through the
slits at a time. Let the line bisecting the line joining the two
slits be the $x$ axis (i.e., $y = 0, x \geq 0$). It is a natural
symmetry axis of the system. After passing through the slits, the
two probability amplitudes propagate with uniform speed $v$ in
spherical waves {\it as a result of diffraction}. In a region in which
these waves do not overlap, the normalized stationary state 
two-particle wavefunction in the $xy$ plane is given by \cite{marchildon}

\begin{equation}
\psi (r_{1A}, r_{2B}, t) = \frac{1}{2 \pi} \frac{e^{i k (r_{1A} +
r_{2B})}}{r_{1A} r_{2B}} e^{\frac{i}{\hbar}E t}\label{eq:wf1}
\end{equation}
where $r_{1A} = \sqrt{x_1^2 + (y_1 - a)^2}$ and $r_{2B} =
\sqrt{x_2^2 + (y_2 + a)^2}$ are the radius vectors of points on
the wave fronts measured from the two slits. This wavefunction is
symmetric under reflection in the $x$ axis together with the
interchange of the particle labels $1\leftrightarrow 2$. The phase
$S (r_{1A}, r_{2B})$ of the wavefunction is

\begin{equation}
S (r_{1A}, r_{2B}, t) = \hbar k (r_{1A} + r_{2B}) - E t
\end{equation}
It is clear from this that the Bohmian trajectories fan out
radially with the slits as the initial positions. Note that a
spherical wavefunction is singular at its origin. Hence, the point
nature of the slits must be understood only in the sense of a {\it limit}.
This is also necessary because otherwise one would get a single
trajectory corresponding to a single initial position rather than
trajectories normal to every point of the spherical wave front,
corresponding to an ensemble of initial positions at the
slit. This is necessary for the compatibility of dBB and SQT for
ensembles. The $x$ and $y$ components of the Bohmian
velocities of the particles are given by

\begin{eqnarray}
v(1)_{x} &=& \frac{1}{m}\frac{\partial S}{\partial
r_{1A}}\frac{\partial r_{1A}}{\partial x_1}
= \frac{\hbar k x_1}{m r_{1A}}\\ v(2)_{x} &=& \frac{1}{m}\frac{\partial
S}{\partial r_{2B}} \frac{\partial r_{2B}}{\partial x_2}|=
\frac{\hbar k x_2}{m r_{2B}}\\ v(1)_{y} &=&
\frac{1}{m} \frac{\partial S}{\partial r_{1A}}\frac{\partial
r_{1A}}{\partial y_1} = \frac{\hbar k (y_1 - a)}{m
y}\\ v_{y_2} &=& \frac{1}{m}\frac{\partial
S}{\partial r_{2B}} \frac{\partial r_{2B}}{\partial y_2} =
-\frac{\hbar k (y_2 + a)}{m r_{2B}}
\end{eqnarray}
Since the spherical waves have the same speed of propagation, we have
$r_{1A}= r_{2B} = vt$, and therefore it follows that

\begin{equation}
v(1)_{x} - v(2)_{x} = \frac{d(x_1 - x_2)}{dt} = \frac{1}{t}
(x_1 - x_2)\label{eq:vx}
\end{equation}
and
\begin{equation}
v(1)_{y} + v(2)_{y} = \frac{d(y_1 + y_2)}{dt} = \frac{1}{t} (y_1 + y_2)
\end{equation}
Solving these equations with the initial conditions $x_1(t_0) - x_2(t_0)
= \delta(0)$ and $y_1(t_0) + y_2(t_0)= \sigma(0)$, one obtains
\begin{eqnarray}
x_1(t) - x_2(t) &=& \delta(0) \frac{t}{t_0}\\
y_1(t) + y_2(t) &=& \sigma(0) \frac{t}{t_0}
\end{eqnarray}
{\it In the limit} $\delta(0) \rightarrow 0$ and $\sigma(0) \rightarrow 0$, we get
\begin{equation}
x_1(t) = x_2(t)\label{eq:x}
\end{equation}
\begin{equation}
y_1(t) = - y_2(t)\label{eq:y}
\end{equation}
at all times $t$. This shows that in this limit the trajectories of the two particles
are at all times symmetrical about the $x$.

If one considers the region where the two spherical waves overlap,
and the particles are bosons, the wavefunction (\ref{eq:wf1}) must
be replaced by

\begin{equation}
\psi (r_{1}, r_{2}, t) = \frac{1}{N}[ \frac{e^{i k (r_{1A} +
r_{2B})}}{r_{1A} r_{2B}} + \frac{e^{i k (r_{1B} + r_{2A})}}{r_{1B}
r_{2A}}]e^{\frac{i}{\hbar}E t}\ \label{eq:wf2}
\end{equation}
where $N$ is a normalization factor, $r_{1B} = \sqrt{x_1^2 + (y_1
+ a)^2}$ and $r_{2A} = \sqrt{x_2^2 + (y_2 - a)^2}$. This is
separately symmetric under reflection in the $x$ axis and the
interchange of the two particles. It follows from the conditions
$r_{1A} = r_{2B} = vt$ and $r_{1B} = r_{2A} = vt$ which must be
satisfied simultaneously that the conditions (\ref{eq:x}) and
(\ref{eq:y})) must still hold. Hence, the Bohmian trajectories of
the two particles are at all times symmetric about the $x$ axis
in this case too. {\it This shows that the trajectories separate into two
disjoint sets, and the theory is non-ergodic}.

Furthermore, the $y$ components of the velocities of the particles
are given by
\begin{eqnarray}
v(1)_{y} &=& \frac{\hbar}{m}{\rm Im} \frac{\partial_{y_1} \psi
(r_{1}, r_{2})}{\psi (r_{1}, r_{2})}\\ v(2)_{y} &=&
\frac{\hbar}{m}{\rm Im} \frac{\partial_{y_2} \psi (r_{1}, r_{2})}
{\psi (r_{1}, r_{2})}
\end{eqnarray}
and therefore
\begin{eqnarray}
v(1)_{y}(x_1(t), y_1(t), x_2(t), y_2(t)) &=& - v(1)_{y} (x_1(t),
-y_1(t), x_2(t), -y_2(t)\\ v_(2){y}(x_1(t), y_1(t), x_2(t),
y_2(t)) &=& - v_(2){y} (x_1(t), -y_1(t), x_2(t), -y_2(t))
\end{eqnarray}
This shows that $v(1)_{y}$ and $v(2)_{y}$ vanish when $y_1(t) =0$
as well as when $y_2(t) =0$, i.e., on the $x$ axis. This implies
that the trajectories of the particles are
not only symmetrical about the $x$ axis, they also do not cross
this axis in the limit $\delta(0) \rightarrow 0$ and $\sigma(0)
\rightarrow 0$. This result is interesting but not essential to show
the incompatibility between dBB and SQT.

The symmetry of the trajectories about the $x$-axis has nontrivial
empirical consequences. If two detectors $D_1$
and $D_2$ are placed at some $x_0$ such that they
are sufficiently asymmetrical about the $x$-axis, the joint detection
probability as a time average will vanish,

\begin{equation}
P^*_{1 2(dBB)} = {\rm lim}_{N \rightarrow \infty} \frac{1}{N}
\sum_{n = 0}^{N - 1} P (\phi_t^n Y)\vert_{D_1, D_2} = 0\label{O}
\end{equation}
where $Y = (y_1, y_2)$ because, by hypothesis, the wavefunction in Bohmian 
theory  acts like a 'pilot wave' that guides the particles via the quantum
potential. It is the particle that can fire a detector but not the 'pilot
wave' itself. On the other hand, the space average is
non-vanishing:

\begin{equation}
\bar{P}_{1 2 (dBB)} = \int_{D_1,\, D_2,\, t} dy_1 d y_2 P
(\,y_1(t), y_2(t)\,) = \bar{P}_{1 2 (SQT)} \neq 0\label{N}
\end{equation}
because the complete ensemble contains correlated pairs with all possible
$y$ values at a given time. This difference is a consequence of the 
non-ergodicity of the Bohmian
motion. Now, SQT predicts (\ref{N}) but not (\ref{O}) in
the limit $\delta(0) \rightarrow 0$ and $\sigma(0)
\rightarrow 0$. It is therefore incompatible with dBB. Any experimental
test of this incompatibility must, of course, take into account the
corrections arising out of nonvanishing $\delta(0)$ and $\sigma(0)$.

Using the formalism of ensembles, one can identify the space mean
as the average over a Gibbs ensemble (the set of identical copies
of the system in all its possible states at a given instant of
time) and the time mean as the average over a time ensemble (the
set of identical copies of the system in different possible states
at different instants of time). The space average can be measured
by using a suitable flux or beam consisting of a sufficiently
large number of identical copies of the system in different states
at the same time, and the time mean by using a sufficiently
attenuated beam so that there is no more than a single copy of the
system in the beam at a given instant of time.

The same conclusions hold for pairs of photons produced by type-I
parametric down-conversion of laser beams in a non-linear crystal.
The pairs are highly correlated in momentum and polarization, and
can be made to pass through a double-slit arrangement, one pair at
a time, and detected by single photon detectors. This type of
experiment is therefore capable of measuring the joint detection
probability of the two photons as a time mean. The appropriate
relativistic quantum mechanical theory that is applicable to
photons is based on the Harish-Chandra-Kemmer formalism
\cite{ghose}. This formalism and the above arguments have been
used to design a realistic experiment with down-converted photon
pairs at Turin \cite{brida}. The Bohmian trajectories of the photons for this type
of experiment have been plotted and will be found in Ref. \cite{ghose1}.
They clearly do not cross the symmetry axis. Importantly, $\sigma(0)$
does not spread at all in this case because there is no velocity dispersion
for photons.

\section{Conclusions}

What we have shown above is generic. One can, in fact, state a
general theorem:

{\it Conventional dBB is incompatible with SQT unless the Bohmian
system corresponding to an SQT system is ergodic}.

I must emphasize one point before concluding. First, the time
average in the examples given above must be measured by joint
detections of the particle positions on identical two-particle
systems prepared successively in time (i.e., over a time ensemble)
and not by repeated measurements on the same system. This is
because the time average of an observable is defined over the
unitary Schr\"odinger evolution of the system (Eqns. (18) and
(36)).

Finally, it is worthwhile drawing attention to what Bohm himself
had to say about the standard interpretation of quantum theory and
his own interpretation\cite{bohm}:
\begin{quote}
``An experimental choice between these two interpretations cannot
be made in a domain in which the present mathematical formulation
of the quantum theory is a good approximation; but such a choice
is conceivable in domains, such as those associated with
dimensions of the order of $10^{- 13}$ cm, where the extrapolation
of the present theory seems to break down and where our suggested
new interpretation can lead to completely different kinds of
predictions."
\end{quote}
The fact that the particular domain referred to by Bohm still
continues to be described very accurately by SQT is irrelevant in
this context. What is significant is that even in domains where
SQT is supposed to be an excellent theory, dBB can be in conflict
with it, and that this difference can only be discovered through
time averages of observables whenever the Bohmian system is
non-ergodic, a feature of his own theory that Bohm seems to have
overlooked.

\section{Acknowledgements}

I am grateful to Anilesh Mohari for many helpful discussions on
ergodicity, and to the Department of Science \& Technology,
Government of India, for a research grant that enabled this work
to be undertaken.

\end{document}